\documentclass[aip,jmp,amsmath,amssymb,reprint]{revtex4-2}

\usepackage{graphicx}
\usepackage{dcolumn}
\usepackage{bm}

\usepackage{graphicx}
\usepackage{dcolumn}
\usepackage{xcolor}
\usepackage{mathtools}

\usepackage{physics}
\usepackage[normalem]{ulem}
\usepackage{lineno}

\usepackage{hyperref} 
\hypersetup{colorlinks=true, citecolor=blue}

\begin{document}

\title{Disentangling superconductor and dielectric microwave losses in sub-micron Nb/TEOS-SiO\textsubscript{2} interconnects using a multi-mode microstrip resonator}

\author{Cougar A. T. Garcia}
\affiliation{Northrop Grumman Corp., Baltimore, MD 21240}
\affiliation{Department of Materials Science and Engineering, University of Maryland, 3121 A. James Clark Hall, College Park, MD 20742, United States}
\author{Nancyjane Bailey}
\affiliation{Northrop Grumman Corp., Baltimore, MD 21240}
\author{Chris Kirby}
\affiliation{Northrop Grumman Corp., Baltimore, MD 21240}
\author{Joshua A. Strong}
\affiliation{Northrop Grumman Corp., Baltimore, MD 21240}
\author{Anna Yu. Herr}
\affiliation{Northrop Grumman Corp., Baltimore, MD 21240}
\affiliation{Present address: IMEC, Leuven, Belgium}
\author{Steven M. Anlage}
\affiliation{Department of Materials Science and Engineering, University of Maryland, 3121 A. James Clark Hall, College Park, MD 20742, United States}
\affiliation{ Quantum Materials Center, Physics Department, University of Maryland, College Park, MD 20742, United States}
\author{Vladimir V. Talanov}
\affiliation{Northrop Grumman Corp., Baltimore, MD 21240}

\date{\today}

\begin{abstract}
Understanding the origins of power loss in superconducting interconnects is essential for the energy efficiency and scalability of superconducting digital logic. At microwave frequencies, power dissipates in both the dielectrics and superconducting wires, and these losses can be of comparable magnitude.
A novel method to accurately disentangle such losses by exploiting their frequency dependence using a multi-mode transmission line resonator, supported by a geometric factor concept and a 3D superconductor finite element method (FEM) modeling, is described. 
Using the method we optimized a planarized fabrication process of reciprocal quantum logic (RQL) 
for the interconnect loss at 4.2 K and GHz frequencies. The interconnects are composed of niobium (Nb) insulated by silicon dioxide made with a tetraethyl orthosilicate precursor (TEOS-SiO\textsubscript{2}). 
Two process generations use damascene fabrication, and the third one uses Cloisonn\'{e} fabrication. For all three, TEOS-SiO\textsubscript{2} exhibits a dielectric loss tangent $\tan\! \delta = 0.0012 \pm 0.0001$, independent of Nb wire width over $0.25 - 4 \: \mu m$.
The Nb loss varies with both the processing and the wire width. For damascene fabrication, scanning transmission electron microscopy (STEM) and energy dispersive X-ray spectroscopy (EDS) reveal that Nb oxide and Nb grain growth orientation increase the loss above the Bardeen–Cooper–Schrieffer (BCS) minimum theoretical resistance $R_{BCS}$. For Cloisonn\'{e} fabrication, the $0.25 \: \mu m$ wide Nb wires exhibit an intrinsic resistance $R_s = 13 \pm 1.4 \: \mu \Omega$ at 10 GHz, which is below $R_{BCS}\approx 17\: \mu \Omega$. That is arguably the lowest resistive loss reported for Nb.
\end{abstract}


\maketitle

\section{Introduction}

Superconducting single flux quantum (SFQ) logic relies on Josephson junctions to form the logic gates, and superconducting transmission lines to deliver clock and power to the gates as well as to propagate bits of information between various circuits on the chip. \cite{Likharev1985, Likharev1991, Hosoya1991, Herr2011, Volkmann2012, Takeuchi2013} The bits are encoded into 
pulses of picosecond duration and millivolt amplitude, which are generated by the junctions\cite{Peterson1977} and each carry a magnetic flux quantum $\Phi _{0} \approx 2.067 \times 10^{-15} \ Wb$.
With the pulse energy of $I_c\Phi_0 \sim 10^{-19} \: J$, where $I_c \sim 100 \: \mu A$ is the junction critical current, and the available clock speed up to 120 GHz,\cite{Yamanashi2010} to compete with CMOS-based computing technologies, the SFQ logic community has been focused on the energy efficiency and scalability. \cite{Tolpygo2016}

Energy efficient SFQ logic families include quantum flux parametron (QFP),\cite{Hosoya1991} reciprocal quantum logic (RQL), \cite{Herr2011} energy efficient single flux quantum (eSFQ) logic,\cite{Volkmann2012} 
and adiabatic quantum flux parametron (AQFP).\cite{Takeuchi2013}
Analogous to Cu/low-k interconnects dominating the net CMOS power budget,\cite{Chang2010power-efficient-computing-technologies} superconducting interconnects can notably impact the energy efficiency of SFQ logic. For instance, AC-powered RQL has been reported to be 300 times more energy efficient than CMOS, including the cooling overhead in large-scale systems.\cite{Herr2011} However, a metamaterial zeroth order resonator (ZOR) delivering clock and power to RQL gates at GHz frequencies dissipated twice more power than the logic junctions, due to the loss in transmission lines forming the ZOR.\cite{Strong2022}

Interconnect density remains one of the main factors limiting the scalability of all forms of superconducting logic.\cite{IRDS2020}
The performance capability of an SFQ logic chip 
increases with the total number of logic gates, which calls for more interconnect layers with smaller wire width and spacing (pitch).
Reducing the wire cross-section to deep sub-micron dimensions, where
it becomes comparable to or smaller than the superconductor magnetic penetration depth $\lambda \sim 90 \: nm$ in Nb, creates complex requirements to fabricate high-bandwidth, low-loss interconnects. Chemical mechanical polishing or planarization (CMP) \cite{Ramzi2012} allows one to fabricate an SFQ logic chip with up to ten wiring layers, \cite{Nagasawa2009} by removing topography left from the previous layer patterning and deposition.\cite{Kaanta1987, Guthrie1992, Krishnan2010}
In our work, both the damascene (metal CMP) and Cloisonn\'{e} (dielectric CMP) planarized processes will be utilized and compared. We will present an elemental material analysis to reveal possible sources of extrinsic microwave loss in the sub-micron Nb interconnects.

The microstrip transmission line (MTL) and stripline are the ubiquitous superconducting interconnects, with 700 GHz analog bandwidth. On one hand, they can provide a basic building block for GHz clock and power distribution systems.\cite{Egan2021, Dai2022, Strong2022} On the other hand, they can form a passive transmission line, to propagate low-bandwidth (35-50 GHz) data with 7-10 SFQ pulses per bit,\cite{Dai2022} or to propagate high-bandwidth (350 GHz) data with single SFQ pulse per bit,\cite{Kautz1978picosecond, Talanov2022} or to produce a delay-line memory. \cite{Tzimpragos2022} The power dissipation is governed by the superconductor and dielectric loss, and is inversely proportional to a transmission line resonator figure of merit, quality factor (Q-factor).\cite{Pozar2011}
Similarly, the propagation distance of SFQ pulse scales with the Q-factor.\cite{Talanov2022} 
Our paper is concerned with simultaneously extracting the superconductor microwave resistance $R_s$ and the insulator loss tangent $\tan\! \delta$ from experimental Q-factor data on MTL resonators, at an RQL operating temperature of liquid helium (LHe) $T=4.2 \: K$. 
\begin{figure*}[t!]
\includegraphics[width=\textwidth]{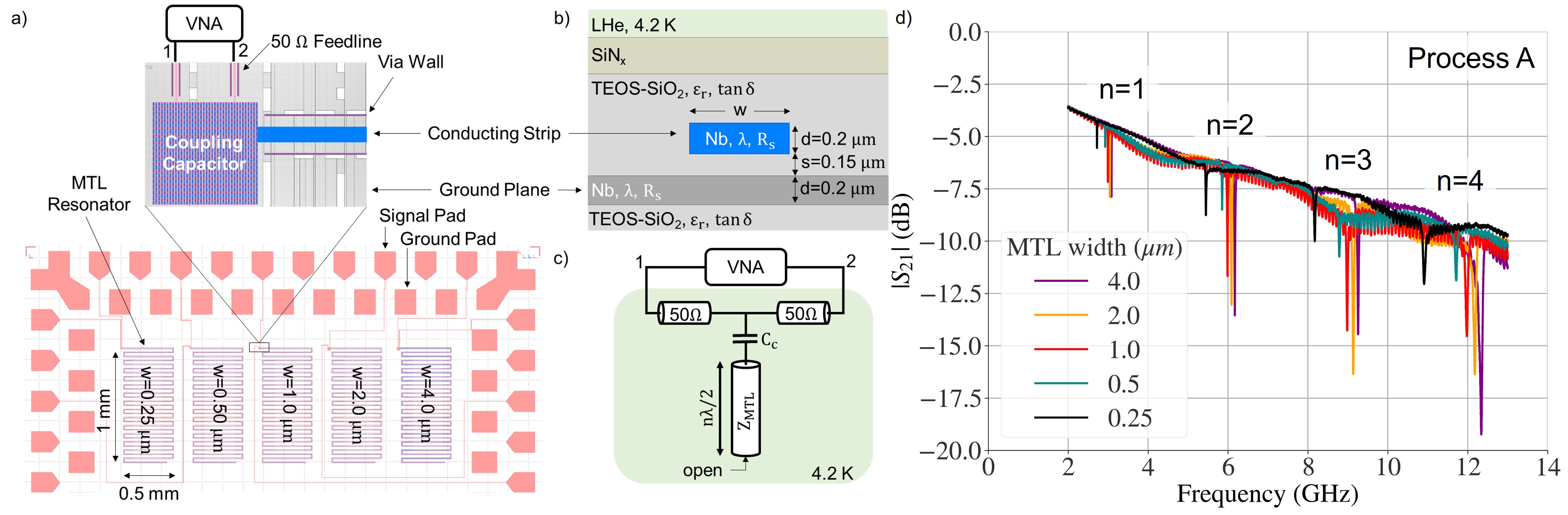}
\caption{
a) Physical layout of a chip with five MTL resonators of conducting strip width $w$ varying from 0.25 to $4 \: \mu m$, fabricated by 3-metal-layer process. Meandering MTL takes up an area $0.5 \times 1 \: mm^2$ per resonator. The zoom-in shows a ``plaid" capacitor, coupling  MTL resonator to $50\ \Omega$ feedline, conceptually connected to vector network analyzer VNA. On the chip, a feedline runs between two sets of ground-signal-ground contact pads.
b) Diagram of the MTL cross-section, showing the nominal thicknesses. Ground plane and conducting strip are embedded into TEOS-SiO\textsubscript{2}, and passivated with SiN\textsubscript{x} layer.
c) Conceptual diagram of the microwave test setup. The resonator is reactively coupled to the $50 \ \Omega$ driving network via a coupling capacitor $C_c$. VNA applies RF power, and measures the transmission coefficient $S_{21}$ at 4.2 K. 
d) Representative $|S_{21}|$ spectra for five MTL resonators from one of the chips fabricated by \emph{Process A}. Resonant dips marked by the mode index $n=1-4$ correspond to the first four $TM_{00n}$ modes. The sloped background is due to the signal attenuation in a cryogenic dip probe.
\label{fig1}}
\end{figure*}

Resonant structures provide the most sensitive way to measure $R_s$\cite{Fairbank1949, Maxwell1949, Turneaure1991, Newman1993, Benvenuti1999, Hein1999, Taber1990, Martens1991, Mazierska1997, Talanov2000, Anl2021} and 
$\tan\! \delta$\cite{Zuccaro1997, Krupka1998, Kaiser2011, Tuckerman2016, Mcrae2020} at microwave frequencies.  We shall employ finite-length sections of MTL, with
open-circuit boundary conditions at the ends,  to act as MTL resonators, enabling sensitive measurements of loss and inductance.\cite{Young1960,Mason1969,Henk1977,Anlage1989,Lang1991} A resonator Q-factor is defined as $Q=2\pi f_r W/P \gg 1$, where $f_r$ is the resonant frequency, $W$ is the energy stored in the resonator, and $P$ is the net power lost by the resonator.
In practice, one measures the loaded Q-factor
\begin{eqnarray}
\frac{1}{Q_l} = \frac{1}{Q_i} + \frac{1}{Q_e} = \frac{1}{Q_c} + \frac{1}{Q_d} + \frac{1}{Q_r} +\frac{1}{Q_e}
\label{Eq.0}
\end{eqnarray}
where $Q_i$ is the internal (unloaded) Q-factor, $Q_e$ is the external (coupling) Q-factor associated with the resonator excitation, and $Q_c$, $Q_d$ and $Q_r$ are the partial Q-factors associated with the conductor, dielectric and radiation power loss, respectively. The coupling contribution $Q_e$ can be removed by modern analysis techniques.\cite{Megrant2012, Khalil2012} The radiation loss\cite{Belohoubek1975} is typically negligibly small and can be ignored. Since $Q_c^{-1}$ and $Q_d^{-1}$ are additive, the corresponding conductor and dielectric loss contributions are inseparable.  To disentangle them, resonant techniques either postulate the ``unwanted" loss, or exploit regimes where $Q_i$ is dominated by the ``wanted" loss. 
The $Q_c \ll Q_d$ regime favors the measurement of $R_s$. 
The $Q_d \ll Q_c$ regime favors the measurement of $\tan\! \delta$. 

Superconducting cavity \cite{Fairbank1949, Maxwell1949, Benvenuti1999, Hein1999} and quasi-optical\cite{Martens1991} resonators measure $R_s$ of bulk and thin-film superconductors by attaining $Q_c \ll Q_d$. To deduce $R_s$ of superconducting thin films using a dielectric resonator technique,\cite{Mazierska1997} one must presume a $\tan \! \delta$ value for the dielectric puck. Tuckerman \emph{et al.} measured $\tan\! \delta$ of Nb/polyimide flexible transmission line tapes in the $Q_d \ll Q_c$ regime,
at 1.2 K and 20 mK where the Nb loss becomes negligible.\cite{Tuckerman2016} Quantum computing resonators attain $Q_d \ll Q_c$ at mK temperatures, to characterize a two-level-system dielectric loss.\cite{Mcrae2020} Kaiser exploited $Q_d \ll Q_c$ in the Nb lumped element resonator, to investigate frequency dependence of $\tan\! \delta$ for amorphous thin-film dielectrics at 4.2 K.\cite{Kaiser2011}
Oates \emph{et al.} reported that at 4 K, losses in Nb/SiO\textsubscript{2} sub-micron stripline resonators are limited by the 
dielectric except for the narrowest strips, although they did not deduce $\tan\! \delta$ or $R_s$.\cite{Oates2017}
Krupka \emph{et al.} optimized a dielectric resonator for $Q_d \ll Q_c$, to measure the loss tangent of isotropic low-loss dielectrics \cite{Krupka1998} and high-resistivity silicon\cite{Krupka2006} at room temperature.
Taber overcame the above limitations by varying the dielectric spacer thickness (geometric factor) of the superconducting parallel-plate resonator,\cite{Taber1990} which allows to deduce both $R_s$ and $\tan\! \delta$.\cite{Taber1990, Talanov2000} However, this approach is impractical for characterization of patterned interconnects.

In contrast to existing work, we shall extract both the superconductor and dielectric loss by exploiting their frequency dependence in a multi-mode  MTL resonator. In fact, we shall take advantage of both losses in our structures being of comparable magnitude $Q_c \sim Q_d$, owing to interplay between the MTL geometry and loss. Fitting theoretical model to experimental dependence of $Q_i^{-1}=Q_c^{-1}+Q_d^{-1}$ on the resonant frequency allows us to quantitatively de-convolve the two losses in a single measurement.

Analytical or numerical modeling of superconducting interconnects\cite{Kautz1978picosecond, Yassin1995, Rafique2005, Belitsky2006, UYen2018, Amini2021} typically relies on the Leontovich boundary condition, commonly referred to as a surface impedance boundary condition (SIBC).\cite{Leontovich1944, Leontovich1948, senior1960impedance, Miller1961}
It approximates the transmitted wave as a wave propagating normal to the surface of an imperfect conductor, regardless of the incident angle. 
SIBC is immensely fruitful for electrically-large systems\cite{Maxwell1949, Klein1990, Turneaure1991, Mazierska1997, Benvenuti1999, Talanov2000, Anl2021} with the conductor surface curvature radius much greater than the penetration depth $\lambda$.
The SIBC approximation is inapplicable to an interconnect with the cross-section comparable to or smaller than $\lambda$.

To overcome this limitation for submicron MTLs, we shall deal the \emph{intrinsic} impedance $Z_s  = R_s +i X_s = \sqrt {i \mu _0 \omega / \sigma }$, where $X_s$ is the microwave reactance, $\mu _0$ is the vacuum permeability, $\omega = 2 \pi f $ is the angular frequency with $f$ being the linear frequency, 
and $ \sigma =  \sigma _1 - i \sigma _2 $ is the superconductor complex conductivity.\cite{Mattis1958} A finite element method (FEM) simulation can provide the field and current distributions within the wire,\cite{Sheen1991, Talanov2022} which for a good superconductor of $\sigma_1 \ll \sigma_2$, or $R_s \ll X_s$, are defined 
by the superfluid electrons distribution, that is $\sigma_2$ or $\lambda$. This allows us to derive a relationship between $Q_c$ and $R_s$ via a geometric factor defined by the MTL cross-section and $\lambda$. Simulating the MTL geometric factor using Ansys' 3D electromagnetic FEM modeler,\cite{ANSYS-HFSS}
we investigate $R_s$ and $\tan\! \delta$ as functions of Nb wire width down to $0.25 \ \mu m$. FEM modeling can also address the fabrication effects impacting interconnects, such as irregular cross-section shape and rounded edges of the wire, critical dimensions miss-targeting and variation across the wafer, intermixing of materials at the interface, etc.

This paper is organized as follows. First, we will describe design, fabrication, and RF characterization of a multi-mode MTL resonator. After presenting data for the frequency dependence of internal Q-factor, we will introduce analytical theory to fit them. Next, a resonator geometric factor, involved into the theory, will be obtained from 3D FEM simulations. Finally, the superconductor and dielectric losses will be deduced, and their dependence on the MTL width and processing conditions will be discussed with the aid of a microscopic analysis.

\section{Resonator design, fabrication, and characterization}

\subsection{Design and layout}

To implement the proposed concept, we designed a chip containing five half-wavelength-long open-ended MTL resonators shown in Fig.~\ref{fig1}(a), representative of RQL interconnects. 
The MTL is formed by a Nb ground plane and a Nb conducting strip, embedded into silicon dioxide derived from tetraethyl orthosilicate (TEOS-SiO\textsubscript{2}), as depicted in Fig.~\ref{fig1}(b).
The conducting strip width $w$ varies from 0.25 to $4 \: \mu m$. 
Each resonator is folded into a meander shape to conserve space. 
Via walls surround the conducting strip, to facilitate RF isolation between the adjacent meander sections.

A superconducting MTL supports a slow $T\!M_{00}$ wave.\cite{Swihart1961, Anl1992} A resonant condition $\beta l_{res} =\pi n$ yields the MTL resonator eigen frequencies
\begin{eqnarray}
f _n = \frac{n}{2 l _{res} \sqrt{L C}} \approx \frac{nc}{2 l_{res} \sqrt{\varepsilon _r}} \sqrt{\frac{s}{s+2 \lambda \coth (d / \lambda )}}  
\label{Eq.1}
\end{eqnarray}
 where $\beta=\omega \sqrt{LC}$ is the phase constant, $l_{res}$ is the resonator geometrical length, 
$n = 1,2,3 . . .$ is the longitudinal mode index, and $L$ and $C$ are the series inductance and shunt capacitance per unit length given in Appendix~\ref{appendix a}, Eqs.~\ref{appendix A eq2b} and \ref{appendix A eq2d}. 
The approximation on the right in Eq.~\ref{Eq.1} holds for a wide MTL (parallel-plate waveguide) with $w \gg s$, where $s$ is the dielectric spacing between the strip and ground plane,\cite{Talanov2000} $c$ is the speed of light in vacuum, $\varepsilon _r$ is the relative dielectric constant, and $d$ is the superconductor wire thickness. The resonator frequency was designed based upon the material nominal properties ($\varepsilon_r=4.2$, $\lambda = 90\: nm$) and fabrication thicknesses ($s=150\ nm$, $d=200\  nm$) shown in Fig.~\ref{fig1}(b). 
The selected resonator length $l_{res}=15 \: mm$ is a trade-off between fitting many resonators on a single $5 \times 5 \: mm^2$ chip, and 
accommodating many resonant modes within the test setup bandwidth. 
With the above parameters, Eq.~\ref{Eq.1} predicts a fundamental frequency $f_1 \approx 3.27 \: \rm GHz$. 
That frequency, measured for a $4 \: \mu m$ resonator, is $f_1=3.20 \pm 0.07 \: \rm GHz$, which is within 0.8\% of the above estimate.
This agreement supports our assumed values of $\varepsilon _r \approx 4.2$ for TEOS-SiO\textsubscript{2},\cite{Rocha2004} and $\lambda \approx 90 \: nm$ for Nb. The latter was verified by SQUID inductance measurements\cite{Luo2021thesis} and agrees with Ref.\cite{Tolpygo2021}

\begin{figure*}[t!]
\includegraphics[width=\textwidth]{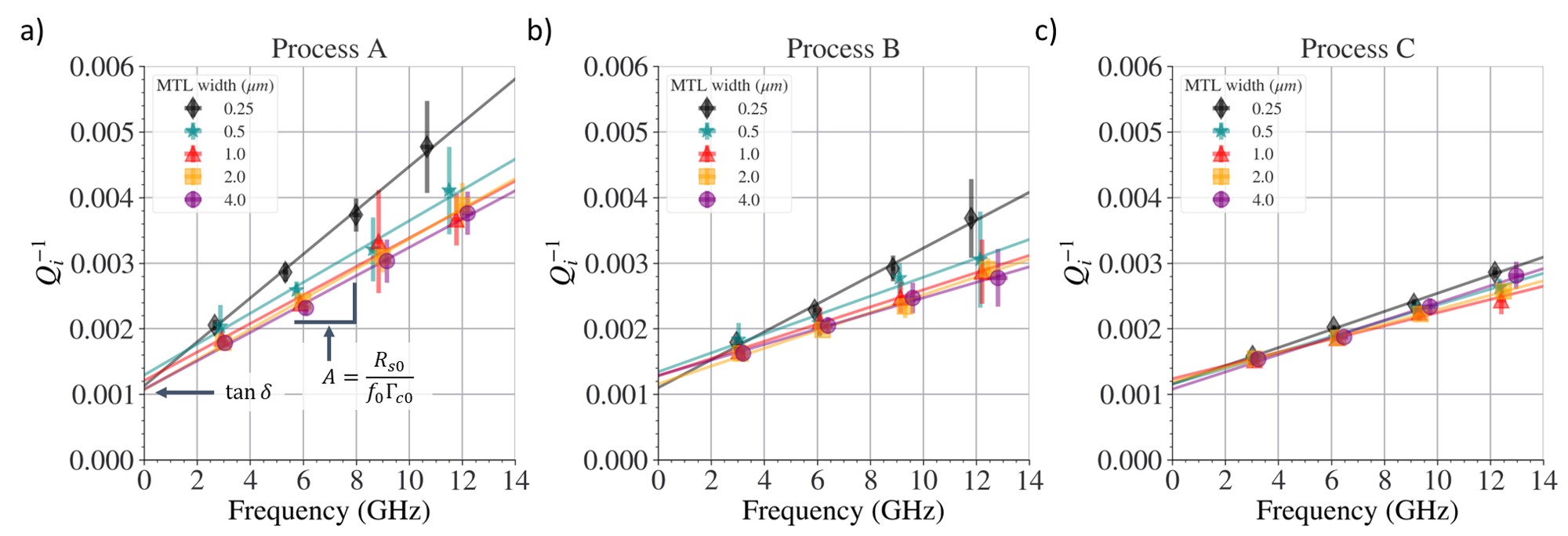}
\caption{
The reciprocal internal Q-factor $Q_i^{-1}$ of MTL resonator at 4.2 K versus the resonant frequencies $f_n$ for the first four modes $n=1-4$. Each graph shows results for all five MTL resonators of varying width $w=0.25-4 \: \mu m$, fabricated by a) \emph{Process A}, b) \emph{Process B}, and c) \emph{Process C}. $Q_i$ was extracted from $S_{21}$ spectra like ones shown in Fig.~\ref{fig1}(d).
The solid lines are linear fits of Eq.~\ref{Eq.8} to data for each MTL width fabricated by respective process. The fit slope $A$ is proportional to the Nb microwave resistance $R_s$ taken at a reference frequency $f_0$, and the fit y-intercept yields the TEOS-SiO\textsubscript{2} dielectric loss tangent $\tan\! \delta$ at GHz frequencies.
\label{fig2}}
\end{figure*}

Each resonator is reactively coupled to a microstrip feedline through a coupling capacitor, as shown in Fig.~\ref{fig1}(a) zoom-in. Figure~\ref{fig1}(c) conceptualizes the RF driving network for our resonators. To overcome the parasitic ripple in transmission coefficient caused by impedance discontinuities in the measurement system (see Fig.~\ref{fig1}(d)), we realized
about $6 \: dB$ insertion loss at the resonance. For a reactively coupled resonator, this corresponds to the critical coupling coefficient $g \equiv Q_i / Q_e = 1$. The network in Fig.~\ref{fig1}(c) can be modeled as a shunt load, with the transmission coefficient $S_{21} = 2 Z_{in} / (Z_0 + 2 Z_{in})$,\cite{Pozar2011} where $Z_{in}$ is the input impedance seen at the T-junction looking toward the coupling capacitor, and $Z_0=50\ \Omega$ is the feedline characteristic impedance. At the resonant frequency, $Z_{in} = R_{in}$ is purely real, and $S_{21} = (1+g)^{-1}$ per Ref.,\cite{Khanna1983} which lead to $2 R_{in} / (Z_0 + 2 R_{in}) = (1+g)^{-1}$. Inserting here the input resistance for a gap-coupled half-wavelength open-ended resonator\cite{Pozar2011} $R_{in}=(8 \pi f_n ^2 C_c ^2 Z_{MTL} Q_i)^{-1}$, and solving the resulting equation for the coupling capacitance $C_c$, gives 
\begin{eqnarray}
C_c = \frac{1}{2 f_n} \sqrt{\frac{g}{\pi Z_0 Z_{MTL} Q_i}}  
\label{Eq.2}
\end{eqnarray}
where $Z_{MTL}=\sqrt{L/C}$ is the MTL characteristic impedance. Inserting $f_1$ given by Eq.~\ref{Eq.1}, $g=1$, $L$ and $C$ given by the parallel-plate approximation Eqs.~\ref{appendix A eq2b}, \ref{appendix A eq2d} in Appendix~\ref{appendix a}, and $Q_i$ of a parallel-plate resonator\cite{Talanov2000} into Eq.~\ref{Eq.2} provides the design value for a coupling capacitor. For our MTL resonators, $C_c$ ranges from 50 to 250 $f\!F$, corresponding to MTL widths ranging from 0.25 to $4 \: \mu m$.

To enable such capacitors in a damascene process, where large metal patches are disallowed because of the requirement to limit the metal coverage density, a ``plaid" capacitor design was used (see Fig.~\ref{fig1}(a), zoom-in). It is formed by multiple parallel wires running in one metal layer, and many more parallel wires running in the adjacent metal layer perpendicular to the wires in the first layer. Two interleaved square grids of vias connect every other wire in the first layer to every other wire in the second layer. This creates two interwoven electrodes with about the same capacitance per unit area as a parallel plate capacitor.

\subsection{Fabrication} 
\label{fabrication}

MTL resonators were fabricated by three generations of a 3-metal-layer process with $0.25\: \mu m$ minimum feature size. Throughout the paper these will be referred to as \emph{Processes A}, \emph{B}, and \emph{C}.
In all three of them, the metal Nb is made by physical vapor deposition, and the insulator SiO\textsubscript{2} is made by low-temperature plasma enhanced chemical vapor deposition \cite{Cote1995} from a tetraethoxysilane Si(OC\textsubscript{2}H\textsubscript{5})\textsubscript{4} (TEOS) precursor. \emph{Process A} is an inverted MTL geometry using a damascene process with the ground plane on the second layer.
\emph{Process B} is a non-inverted MTL geometry using a damascene process with the ground plane on the first layer.
\emph{Process C} is a non-inverted MTL geometry using a Cloisonn\'{e} process with the ground plane on the first layer.
Figure~\ref{fig1}(b) depicts the cross-sectional geometry of the MTL fabricated by \emph{Processes B} or \emph{C}. 

The damascene process begins with depositing a uniform layer of dielectric on a planarized surface.
After the trenches or vias corresponding to interconnects are defined by photolithography and subtractively patterned 
using reactive-ion etching (RIE), metal is deposited to fill and overfill (overburden) the trenches or vias. Finally, CMP polishes away the excess metal until the intermetal dielectric, embedding the wires or vias, is exposed. To promote adhesion of TEOS-SiO\textsubscript{2} to Nb, the planarized surface is treated with an oxygen plasma. The smooth surface is now ready for the next layer. 

The Cloisonn\'{e} process begins with depositing a uniform layer of interconnect metal on a planarized surface. 
After the interconnects are defined by photolithography and subtractively patterned 
using RIE, the dielectric is deposited conformally over the metal features.
Once metal is completely embedded with dielectric, the resulting surface is planarized by CMP to remove the excess dielectric, until the wires or vias are exposed. The surface is now ready for the next layer.

\subsection{Experimental}

From each process, 6 to 9 chips selected within the inner 80 mm diameter of the 150 mm wafer were tested.
Measurements were taken at $4.2 \: K$ in a LHe Dewar using an RF dip probe equipped with a 32-contact-pad test fixture.\cite{Dai2022, Egan2021, Talanov2022, Strong2022} 
To provide for fast sample exchange, a flip-chip press-contact technology is used, where the chip contact pads are pressed against the fixture non-magnetic Cu/Au bumps. The fixture PCB interfaces the bumps to the probe semi-rigid coaxial cables.
During experiment, the fixture and roughly $30 \: cm$ of the dip probe were immersed in a LHe bath. The S-parameters were measured by a Keysight N5222A 2-port vector network analyzer (VNA), with the test cables calibrated to the top of the probe. To minimize the possibility of non-linear effects\cite{Chin1992, Golos1995, Jutzi2003} in MTL resonators, the microwave power entering the chip was kept below $10\: \mu W$. 

Figure~\ref{fig1}(d) shows representative transmission coefficient magnitude $|S_{21}|$ vs. frequency for one of the chips made by \emph{Process A}. The probe bandwidth of 14 GHz limited the measurements to the first four modes $n=1-4$ of the MTL resonators. To extract the internal Q-factor $Q_{i}$ and the resonant frequency $f_n$ from the complex $S_{21}$ data, we subtract a phase delay due to the probe cables and employ a diameter correction method.\cite{Peter1998, Khalil2012, Megrant2012} Depending on the fabrication process, MTL geometry, and mode index, we observed $Q_{i}$ between 200 and 700.

Figure~\ref{fig2} shows $Q_{i}^{-1}$ obtained from 4 wafers on a total of 22 chips (5 MTL resonators per chip). Each panel of Fig.~\ref{fig2} shows $Q_i^{-1}$ as a function of resonant frequency for all five MTL widths, for particular fabrication process. Each data point and error bar represent the arithmetic mean and 1 standard deviation for a sample of 6-9 chips per process, to statistically describe the loss variation for each respective MTL width for each process. The increased error bars for modes $n=3,4$ in \emph{Processes A} and \emph{B} are due to parasitic ripple in $|S_{21}|$ seen in Fig.~\ref{fig1}(d). Evaluating a measurement repeatability by re-inserting the same chip several times into a test fixture, yielded maximum spread in the extracted $Q_i$ and $f_n$ of $<\!10\%$ and $< \!0.1\%$, respectively.

\section{Data Analysis}

To explain the nearly linear dependence of $Q_i^{-1}$ on resonant frequency observed in Fig.~\ref{fig2},
a geometric factor concept devised in Appendix~\ref{appendix a} allows to express $Q_i$ as
\begin{eqnarray}
\frac{1}{Q_i} =\frac{1}{Q_c}+\frac{1}{Q_d}= \frac{R_{s}^{gp}}{\Gamma_c^{gp}} + \frac{R_{s}^{cs}}{\Gamma_c^{cs}} + \frac{\tan \! \delta}{\Gamma_d}
\label{EqQiGamma}
\end{eqnarray}
where $R_{s}^{gp}$ and $R_{s}^{cs}$ are the intrinsic resistances of the ground plane and conducting strip, $\Gamma_c^{gp}$ and $\Gamma_c^{cs}$ are the partial geometric factors associated with resistive loss in the ground plane and conducting strip, $\tan \! \delta$ is the TEOS-SiO\textsubscript{2} dielectric loss tangent, and $\Gamma_d$ is the geometric factor associated with loss in the TEOS-SiO\textsubscript{2}.
By making assumptions about the frequency dependence of $R_s$ and $\tan\! \delta$, we shall factor out the frequency in Eq.~\ref{EqQiGamma}. 

A frequency dependence for $R_s$ can be written as
\begin{eqnarray}
R_s = R_{s0} (\omega / \omega _0)^\alpha
\label{eq5}
\end{eqnarray}
where $R_{s0}=R_s(\omega_0)$ is the intrinsic resistance at a reference frequency $\omega _0 = 2\pi f_0$ that we choose $f_0 = 10 \ \rm GHz$. Although a BCS theory predicts that the scaling exponent $\alpha$ decreases from 1.8 to 1.7 over 1--10 GHz for Nb at 4.2 K, the experiments show that the smearing density of states makes $\alpha \approx 2$ up to 10 GHz,\cite{Philipp1983, Turneaure1991} which roughly corresponds to the frequency span of our resonators. For an isotropic good superconductor with $\sigma _1 \ll \sigma _2$, the two-fluid model also predicts $\alpha=2$.\cite{Newman1993, Turneaure1991} 

A dielectric loss tangent can be expressed as $\tan\! \delta = \sigma _d / \varepsilon _0 \varepsilon _r \omega$, where $\sigma _d$ is the material conductivity, and $\varepsilon_0$ is the vacuum permittivity.\cite{Pozar2011} According to Jonscher’s universal relaxation law, $\sigma _d$ scales with frequency as $\sigma _d = \sigma _{DC} + P \omega ^p $,  where $\sigma _{DC}$ is the DC conductivity, $P$ is the exponential prefactor, and the exponent $p$ falls in the range of $0 < p \le 1$.\cite{Jonscher1977} Then, at frequencies high enough that $\sigma _{DC} \ll P \omega ^ p$, a frequency dependence for $\tan\! \delta$ can be written as
\begin{eqnarray}
\tan\! \delta = {\tan \! \delta}_0 (\omega / \omega _0)^{p-1} 
\label{tand}
\end{eqnarray}
 where $\tan\! \delta _0 = P \omega _0 ^{p-1} / \varepsilon_0 \varepsilon_r$ is the loss tangent at a reference frequency $\omega _0$. Jonscher proposed that low-loss dielectrics with $\tan\! \delta < 0.1$ have a nearly ``flat loss", that is $p \rightarrow 1$, over several decades of frequency.\cite{Jonscher1977} However, $\tan\! \delta$ can also increase approximately linearly with frequency in some ceramics, glasses and polymers,\cite{Zuccaro1997, NIST-Baker-Jarvis2012} corresponding to $p \approx 2$ in Eq.~\ref{tand}. 

Recalling that $\Gamma_c \propto \omega$ per Eq.~\ref{appendix A eq4a}, application of Eqs.~\ref{eq5} and \ref{tand} to Eq.~\ref{EqQiGamma} gives 
\begin{eqnarray}
\frac{1}{Q_i}= \left( \frac{R_{s0}^{gp}}{\Gamma_{c0}^{gp}} + \frac{R_{s0}^{cs}}{\Gamma_{c0}^{cs}} \right) \left(\frac{\omega}{ \omega_0}\right)^{\alpha-1} \!+ \frac{{\tan \!\delta}_0}{\Gamma_d} \left(\frac{\omega}{\omega _0}\right)^{p-1}
\label{Eq.7}
\end{eqnarray}
where the quantities with subscript ``0" are taken at a reference frequency $\omega_0$. 
Considering $\alpha=2$ for Nb, the linear dependence of $Q_i^{-1}$ with frequency in Fig.~\ref{fig2} 
suggests $p \approx 1$ in Eq.~\ref{Eq.7}. This implies that for all three processes TEOS-SiO\textsubscript{2} has a ``flat loss" over at least 3-13 GHz. That result agrees with Tuckerman \emph{et al.},\cite{Tuckerman2016} who observed similar behavior in a Nb/polyimide transmission line resonator at 4.2 K over 2-20 GHz. However, Kaiser\cite{Kaiser2011} reported $0.57 \le p \le 0.68$ for various amorphous thin-film dielectrics at 4.2 K and 0.1-20 GHz, although he ignored the Nb loss in a lumped element resonator.

An embedded MTL like in Fig.~\ref{fig1}(b) has $\Gamma_d \approx 1$. Then, setting $\alpha=2$ and $p=1$ in Eq.~\ref{Eq.7} leads to the linear form
\begin{eqnarray}
Q_i ^{-1} = \frac{R_{s0}}{ f_0 \Gamma_{c0}} f + \tan\! \delta = A f + \tan\! \delta
\label{Eq.8}
\end{eqnarray}
where $R_{s0}=(R_{s0}^{gp}\Gamma_{c0}^{cs}+R_{s0}^{cs}\Gamma_{c0}^{gp})/(\Gamma_{c0}^{gp} + \Gamma_{c0}^{cs})$ represents the MTL net resistive loss, and  $\Gamma _{c0} = \left(1/\Gamma_{c0}^{gp} + 1/\Gamma_{c0}^{cs}\right)^{-1}$ is the net conductor geometric factor. $R_{s0}=R_{s0}^{cs}=R_{s0}^{gp}$ in the case of the homogeneous MTL.
Fitting Eq.~\ref{Eq.8} to the data sets in Fig.~\ref{fig2} with $A=R_{s0}/f_0 \Gamma_{c0}$ and $\tan\! \delta$ being the fitting parameters, 
the fit slope $A$ is proportional to $R_{s0}$, and the fit y-intercept is $\tan\! \delta$.

Figure~\ref{fig2} shows that {\emph{Process A}} exhibits larger fit slope $A$, hence higher superconductor loss, for all MTL widths relative to {\emph{Processes B}} and \emph{C}. 
For MTL width $1 \: \mu m$ and under, {\emph{Process B}} has higher superconductor loss in comparison to {\emph{Process C}}, 
while for the $2$- and $4$-$\mu m$-wide  MTLs {\emph{Processes B}} and {\emph{C}} yield similar loss. Furthermore, all MTL widths and processes yield roughly the same y-intercept, corresponding to $\tan\! \delta \approx 0.0012$.

Equation~\ref{Eq.8} infers that the MTL superconductor loss can be deduced from the fit slope $A$ as 
\begin{eqnarray}
R_{s0} = A f_0 \Gamma _{c0}
%
\label{RsfromA}
\end{eqnarray}
For the wide MTL with $w\gg s$, the conductor geometric factor can be found from a parallel-plate model\cite{Talanov2000}
\begin{eqnarray}
\Gamma_{c0}^{PP}=\omega _0 \mu _0 \frac{s + 2 \lambda \coth (d/\lambda)}{2 [\coth (d/\lambda) + (d/\lambda) \csch ^{2} (d/\lambda)]}
\label{Eq.4b}
\end{eqnarray}
Finding $\Gamma_c$ for the narrow MTL with $w \lesssim s$ calls for FEM modeling because the parallel-plate approximation accounts for neither the current concentration at the edges of conducting strip\cite{Sheen1991} nor the field fringing\cite{Chang1979} nor the case of irregular geometry of the conducting strip.\cite{Guo2020}

Equation~\ref{Eq.0} infers that a resonator optimized for simultaneous characterization of both superconductor and dielectric losses calls for $Q_c=Q_d$, which according to Eq.~\ref{Eq.8} corresponds to a conductor geometric factor 
$\Gamma_{c0}^*= R_{s0}/ \tan\! \delta$. For representative $R_{s0} =20\: \mu \Omega$ and $\tan\! \delta = 10^{-3}$,
one estimates $\Gamma_{c0}^* = 20\: m\Omega$. For the MTL geometry depicted in Fig.~\ref{fig1}(b), Eq.~\ref{Eq.4b}
predicts $\Gamma_{c0}^{PP} \sim 12\: m\Omega$ at 10 GHz and 4.2 K, which is close to $\Gamma_{c0}^*$. This makes our MTL resonators sensitive to both types of loss.

\section{FEM modeling}
\label{section:HFSS}

In our work, the conductor geometric factor $\Gamma_c$ depends on the field and current distributions inside the wires (see Eqn.~\ref{appendix A eq4a}).
Arguably, Ref.~\cite{Sheen1991} is the only paper solving these for a superconducting strip transmission line, using a proprietary FEM solver.
We employ Ansys' High Frequency Structure Simulator (HFSS)\cite{ANSYS-HFSS} to find geometric factors of the MTL resonators.

HFSS is a full 3D FEM simulator, capable of solving fields, currents, and network parameters for virtually any microwave structure with  the ratio of largest to smallest dimension less than $10^4$. 
The peculiar aspects of modeling a superconductor network in HFSS are
(i) defining the lossy superconductor material by a real conductivity and a negative permittivity; \cite{Mei1991}
(ii) using perfect electric conductor (PEC) material to connect superconductor members to the ports, since HFSS does not allow superconducting ports;\footnote{At the time of this work, the most recent version of HFSS that permits material with complex conductivity and removes a need for PEC ports, was unavailable}
(iii) enforcing the \emph{solve-inside} option for the superconductor material, to override the HFSS default setting of only solving inside a material when the conductivity is less than $10^5 \: S/m$.
\begin{figure*}[t!]
\includegraphics[width=\textwidth]{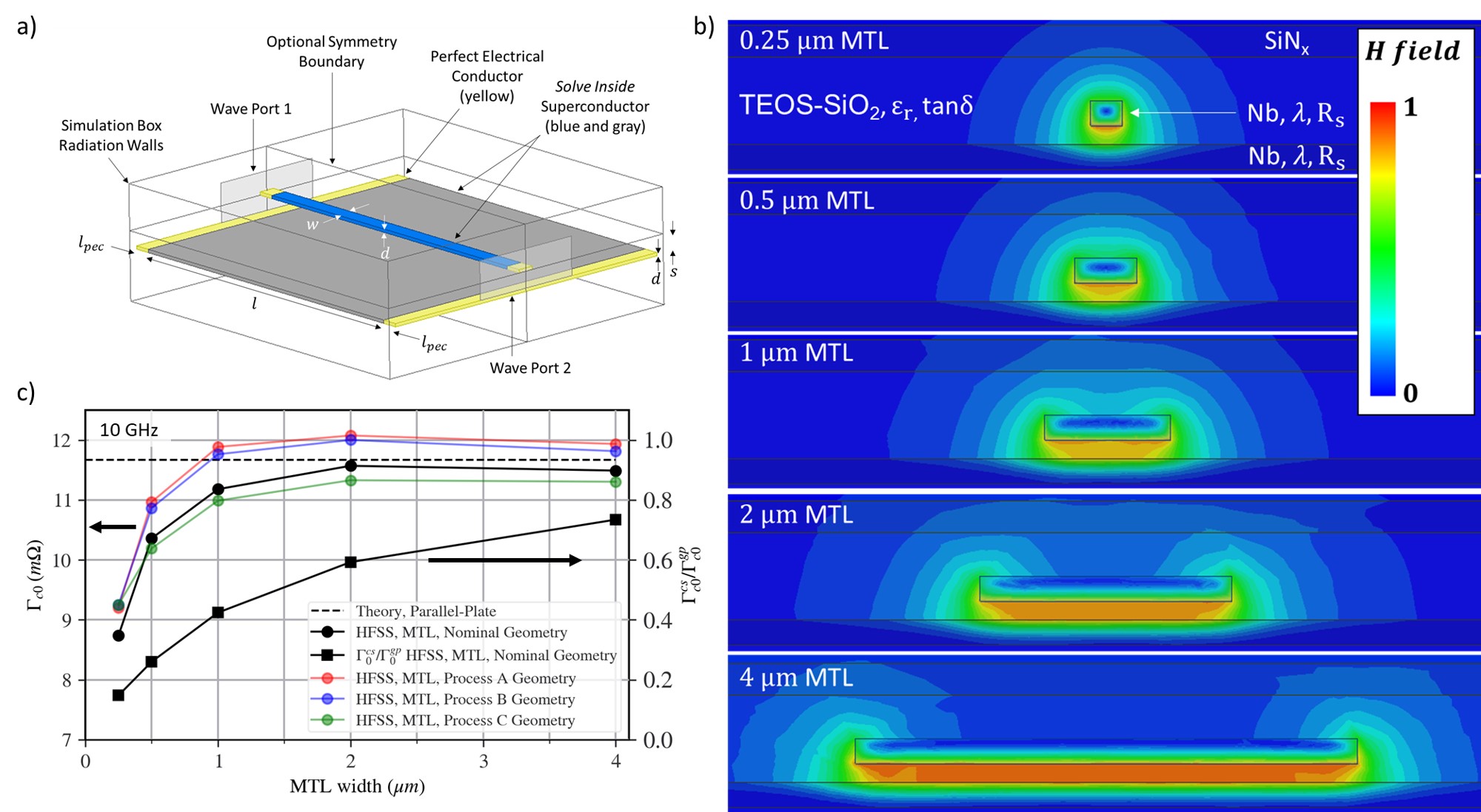}
\caption{  
a) The HFSS model of a superconducting MTL. A conducting strip (blue) of width $w$ is separated by a dielectric spacing (transparent) of thickness $s$ from a ground plane (gray). 
Both the strip and ground plane have same thickness $d$ and simulation length $l$. 
The MTL is enclosed by a simulation box with all faces set as \emph{Radiation} boundaries. 
To stimulate the model, PEC members (yellow) connect the wave ports to the superconductor ground plane and conducting strip.
b) Magnetic field intensity looking into MTL cross-section, simulated at 10 GHz. The MTL width varies from $0.25$ to $4 \: \mu m$, top to bottom. Here, $s=150 \: nm$, $d=200 \: nm$, the Nb magnetic penetration depth and intrinsic resistance are $\lambda = 90 \: nm$ and $R_s = 20 \: \mu \Omega$, the TEOS-SiO\textsubscript{2} relative dielectric constant and loss tangent are 4.2 and $10 ^{-3}$, and the SiN\textsubscript{x} relative dielectric constant and loss tangent are 7.5 and $10 ^{-4}$.
c) The net conductor geometric factor $\Gamma _{c}$ taken at 10 GHz versus MTL width, found from HFSS solver using Eq.~\ref{HFSS: Eq. gamma_c0} for the nominal cross-sectional geometry (black circles),
and for the fabricated geometries for all three processes (red, green, and blue circles).
Dashed black line is a parallel-plate approximation using Eq.~\ref{Eq.4b}.
The black squares show the geometric factor ratio $\Gamma_{c0}^{cs}/\Gamma_{c0}^{gp}$.
Solid lines are a guide to the eye.
\label{fig3}}
\end{figure*} 
\subsection{Defining a superconductor in HFSS}

Inserting the superconductor current-field constitutive relationship $\textbf J = (\sigma _1 - i \sigma _2) \textbf E$ into Maxwell’s curl-H equation yields
\begin{eqnarray}
\nonumber
\nabla \times \textbf H = \sigma _1 \textbf E + i \omega \varepsilon _0 \left(\varepsilon - \frac{\sigma _2}{\varepsilon _0 \omega}\right) \textbf E    
\label{theory: Eq.12}
\end{eqnarray}
where the parenthesis enclose the superconductor relative permittivity $\varepsilon _{sc} = \varepsilon - \frac{\sigma _2}{ \varepsilon _0 \omega}$, 
with $\varepsilon$ being the ordinary dielectric constant associated with displacement current. 
Modeling a superconductor as a collisionless neutral plasma \cite{Mei1991, Mishonov1991} with Langmuir frequency $\omega _{p} = c / \lambda $ 
and dielectric function $\varepsilon _p = 1 - (\frac{\omega _{p}}{\omega})^2 = 1 - (\frac{c}{\omega \lambda})^{2}$, comparison of $\varepsilon _{sc}$ and $\varepsilon _p$ infers that $\varepsilon =1$ is the superconductor permittivity in the limit $\omega \gg \omega _p$,~\footnote{For niobium, $\omega _p$ is above the superconducting gap frequency} and $\sigma _2 = 1/\omega \mu _0 \lambda ^2$ is the London conductivity.
Therefore, a lossy superconductor can be defined as a material with real conductivity $\sigma_{sc} = \sigma_{1}$ and real permittivity $\varepsilon_{sc} = 1- \frac{\sigma_2}{\varepsilon_0 \omega}$, applicable over the entire range of temperatures and frequencies of interest. Owing to $\sigma _2 \gg \varepsilon _0 \omega$, a superconductor permittivity is substantially negative quantity. \cite{Glover1957}

Generally, the Mattis-Bardin theory allows to tabulate $\sigma_{sc}$ and $\varepsilon_{sc}$ as HFSS \emph{datasets}. \cite{Talanov2022} Here, exploiting the assumption $\sigma _1 \ll \sigma _2$, we define a superconductor in HFSS as a frequency-dependent material. 
Relating $\sigma_1$ to the experimental intrinsic resistance and penetration depth, a two-fluid model gives $\sigma_{sc}^{HFSS} = 2 R_{s0}/ \omega _0 ^{2} \mu _0 ^{2} \lambda ^{3}$ that is frequency independent. 
By the same assumption, modeling $\sigma _2$ as the London conductivity gives $\varepsilon_{sc}^{HFSS} = - (c/\omega \lambda)^{2}$. 
Adapting HFSS to solve \emph{inside} the superconductor overcomes a limitation of SIBC-based models\cite{Yassin1995, Rafique2005, Belitsky2006, UYen2018, Amini2021} to an interconnect with a much bigger cross-section than $\lambda$.

\subsection{Modeling a superconducting MTL in HFSS}

Figure~\ref{fig3}(a) shows 3D HFSS model of a two-port network formed by a superconductor MTL with PEC ports of
the same cross-sectional geometry. \cite{Talanov2022}
Note that to determine $\Gamma_c$, we simply model a piece of uniform MTL, not an MTL resonator.
To achieve accurate results, the model is electrically short $\beta l \ll 1$, with $l$ being the length of the MTL piece, and the PEC port length is a small fraction of the simulated transmission line $l_{pec} \ll l$. 
The \emph{network analysis driven terminal} solution type \cite{ANSYS-HFSS} was used to simulate MTL lengths $l = 10-50\: \mu m$ with PEC port length  $l_{pec} = 1 \: \mu m$, at simulation frequency $10 \: \rm{GHz}$. PEC ports were de-embedded to get only the network parameters pertaining to the superconductor MTL.

All geometries and material definitions were parameterized and the \emph{Optimetrics} option was used for parameter sweeps. 
To provide for fast and accurate simulation, the HFSS convergence criteria was set at 0.1-1\% for the $RLGC$ 
parameters defined as $R^{HFSS}=\Re(Z)$, $L^{HFSS}=\Im(Z)/\omega$, $G^{HFSS}=\Re(Y)$, and $C^{HFSS}=\Im(Y)/\omega$.\cite{CATG2022} Here, $Z$ and $Y$ are the series impedance and shunt admittance of a general transmission line,\cite{Ramo1994} given by
\begin{subequations}
\label{EqsZY}
\begin{gather}
Z = 2 (Z_{11} - Z_{12}) l^{-1} \label{appendix B eq3a} \\[1ex]
Y = Z_{12}^{-1} l^{-1} \label{appendix B eq3b}
\end{gather}
\end{subequations}
where $Z_{11}$ and $Z_{12}$ are the elements of the network [Z]-matrix found by HFSS, as described in Appendix~\ref{appendix b}. It is tractable to complete parametric sweeps on the order of 100 simulations in a few hours.

Figure~\ref{fig3}(b) shows the magnetic field intensity looking into MTL cross-section, for all five MTL widths, in the case of the nominal geometry and material parameters for \emph{Processes B} and \emph{C}. For $2 \: \mu m$ wide MTL and above, the magnetic field distribution has a parallel-plate-like geometry. For $1 \: \mu m$ MTL and below, the magnetic field penetrates the majority of the conducting strip, but only a small fraction of the ground plane.

\subsection{Deducing geometric factor from HFSS}

Setting the dielectric loss in the HFSS model to zero leads to $Q_i = Q_c$. 
The net conductor geometric factor can be found from HFSS simulations, ran at a reference frequency $\omega_0$, as
\begin{eqnarray}
\Gamma _{c0} = R_{s0}^{HFSS} \frac{\omega_0 L^{HFSS}}{R^{HFSS}}
\label{HFSS: Eq. gamma_c0}
\end{eqnarray}
where $R_{s0}^{HFSS}$ is the intrinsic microwave resistance corresponding to $\sigma_{sc}^{HFSS}$.
Likewise, the partial geometric factor of just the conducting strip can be found by setting the ground plane loss to zero in the above procedure, and vice versa for the ground plane geometric factor. For each MTL geometry, only a single simulation for any resistance value meeting 
$R_{s0} ^{HFSS} \ll \omega_0 \mu _0 \lambda $ is required,
which substantially accelerates the data analysis.
We hypothesize that geometric factor can also be computed by numerically integrating Eq.~\ref{appendix A eq4a} with the fields and currents found by an FEM simulator (see Fig.~\ref{fig3}(b)). 

The geometric factors found using HFSS were verified to be independent of 
$R_{s0}^{HFSS}$ between 0.1 and $100 \: \mu \Omega$. The HFSS results were further validated by comparing the geometric factor by Eq.~\ref{HFSS: Eq. gamma_c0} to the parallel-plate approximation Eq. \ref{Eq.4b}.
Figure~\ref{fig3}(c) shows that $\Gamma _{c}$ for the MTL with nominal geometry (black solid line) approaches $\Gamma _{c}^{PP}$ (black dashed line) within 1.5\% at $2 \: \mu m$ MTL width and above. 
Intuitively, this can be understood from Figure~\ref{fig3}(b), where the fringing fields above the conducting strip begin to overlap below a $2 \: \mu m$ width. 
Hence, Eq.~\ref{Eq.4b} can be used for preliminary data analysis in wide MTLs with $w⁄s > 13$, without resorting to HFSS simulations. 

\section{Results and Discussion}

The cross-sectional geometries for each \emph{Process} and MTL width were measured by focused ion beam (FIB) or scanning transmission electron microscopy (STEM). These actual geometries, including slanted sidewalls of conducting strip, were modeled in HFSS to find the conductor geometric factor $\Gamma_{c0}$ using Eq.~\ref{HFSS: Eq. gamma_c0}.
Finally, such actual geometric factors shown in Fig.~\ref{fig3}(c) were utilized to deduce the MTL net intrinsic resistance $R_{s0}$ from the linear fits in Fig.~\ref{fig2} using Eq.~\ref{RsfromA}. 

\begin{figure}[t!]
\includegraphics[width=.48\textwidth]{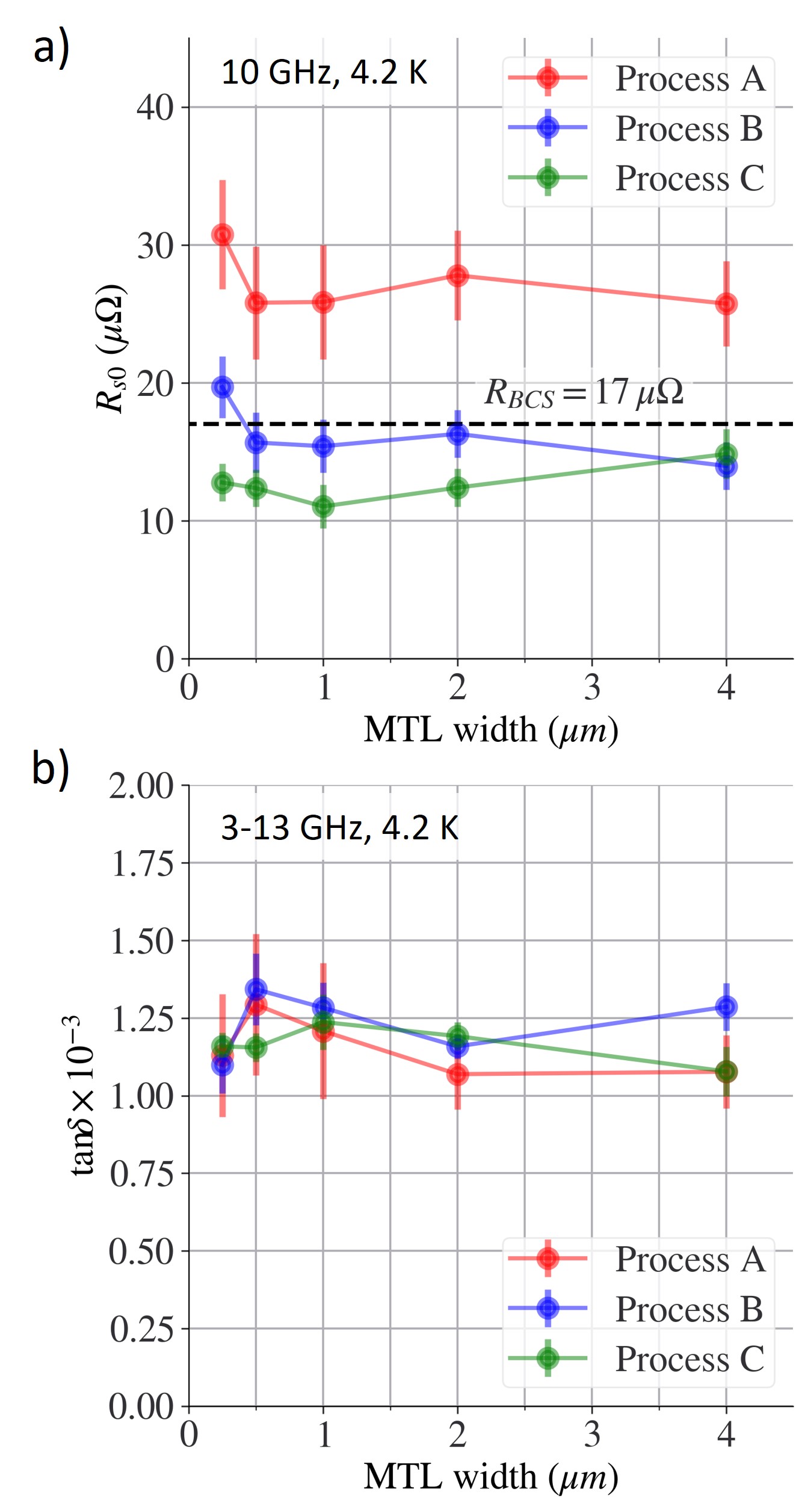}
\caption{
(a) The MTL net intrinsic resistance at 10 GHz, 4.2 K versus MTL width and fabrication process, deduced from the fit slope $A$ in Fig.~\ref{fig2} using Eq.~\ref{RsfromA} and the geometric factor $\Gamma _{c0}$ given by Eq.~\ref{HFSS: Eq. gamma_c0}. 
The black dashed line represents the BCS theoretical minimum for Nb. \cite{SRIMP, Benvenuti1999}
(b) TEOS-SiO\textsubscript{2} loss tangent $\tan\! \delta$ at 4.2 K versus MTL width and fabrication process given by the fit y-intercept in Fig.~\ref{fig2}.
Solid lines are a guide to the eye. 
\label{fig4}}
\end{figure}

Figure~\ref{fig4} shows the results vs MTL width. The error bars in Fig.~\ref{fig4}(a) are obtained by applying the error propagation analysis to Eq.~\ref{RsfromA}, neglecting correlations. The uncertainties in the fit slope $A$ and the y-intercept $\tan\! \delta$ are estimated from the linear regression in Fig.~\ref{fig2}. The uncertainty in $\Gamma _{c}$ is found using HFSS to model the effects of MTL geometry variations across the wafer. \cite{CATG2022}

Since the ratio $\Gamma_{c0}^{cs}/\Gamma_{c0}^{gp} \le 1$ decreases with MTL width per Fig.~\ref{fig3}(c), the smaller the width, the greater the conducting strip contribution into ${R_{s0}}$. In the case of dissimilar $R_s^{cs} \neq R_s^{gp}$, the ${R_{s0}}$ in Eq.~\ref{Eq.8} can be written as
\begin{eqnarray}
R_{s0}= 
R_{s0}^{cs}\frac{1 + (\Gamma_{c0}^{cs}/\Gamma_{c0}^{gp})(R_{s0}^{gp}/{R_{s0}^{cs})}}{1+\Gamma_{c0}^{cs}/\Gamma_{c0}^{gp}} \nonumber\\ \approx R_{s0}^{cs}\left[1-\frac{\Gamma_{c0}^{cs}}{\Gamma_{c0}^{gp}}\left(1-\frac{R_{s0}^{gp}}{R_{s0}^{cs}}\right)\right]
\label{Eq. Qc}
\end{eqnarray}
where the approximation holds for $\Gamma_{c0}^{cs}/\Gamma_{c0}^{gp} \ll 1$.
For the $0.25 \: \mu m$ wide MTL, assuming $R_{s0}^{gp} \sim R_{s0}^{cs}$, the ratio $\Gamma_{c0}^{cs}/\Gamma_{c0}^{gp} \sim 0.14$ means that the corresponding $R_{s0}$ in Fig.~\ref{fig4}(a) is dominated by the strip loss.

Figure~\ref{fig4}(a) reveals that for all MTL widths, the Nb loss reduces from \emph{Process A} to \emph{Process B} to \emph{Process C}. Our medium-loss \emph{Process B} has $R_{s0}$ in the $14-20\: \mu\Omega$ range. This is in agreement with $R_{s0} \approx 16\: \mu\Omega$ measured by the parallel-plate resonator for Nb thin films at 4.2K, 12 GHz \cite{Taber1990} and scaled to 10 GHz using Eq.~\ref{eq5}. Our result is also in agreement with the Bardeen–Cooper–Schrieffer (BCS) minimum intrinsic resistance $R_{BCS} \approx 17 \: \mu\Omega$ measured by Benvenuti \emph{et al.} using the RF cavity with thin-film Nb walls at 4.2 K, 1.5 GHz \cite{Benvenuti1999} and scaled to 10 GHz, which is shown by the black dashed line in Fig.~\ref{fig4}(a). Our low-loss \emph{Process C} shows that sub-micron Nb interconnects can, remarkably, have $R_{s}$ below $R_{BCS}$ at 4.2 K, which can be attributed to the smearing of density of states.\cite{Philipp1983}
Thus, \emph{Processes B} and \emph{C} demonstrate that a CMP planarized Nb interconnect can be scaled into submicron dimensions with no penalty above the minimum theoretical loss. 
\begin{figure*}
\includegraphics[width=\textwidth]{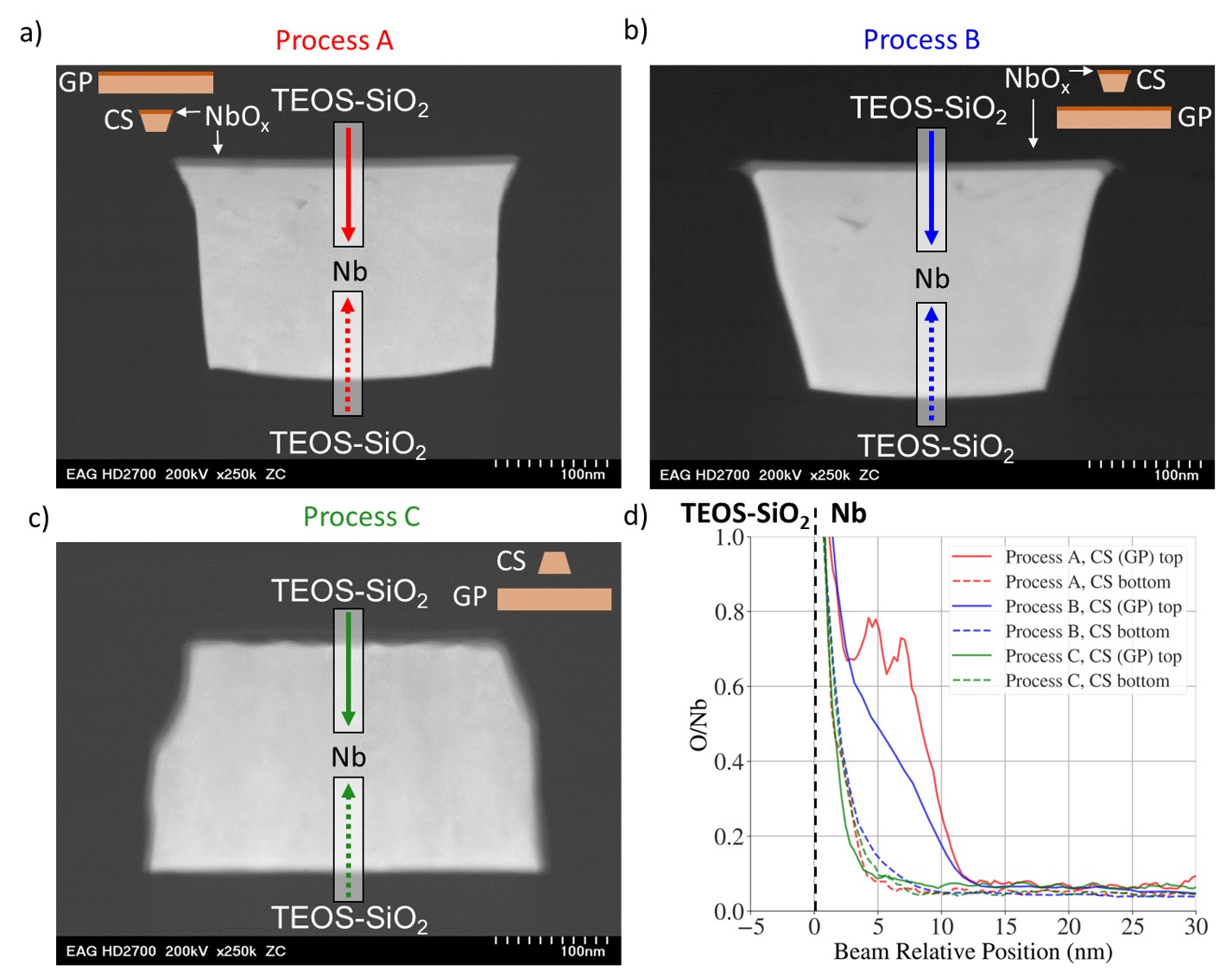}
\caption{STEM cross-sections of conducting strip for representative $0.25 \: \mu m$ MTL resonators fabricated by a) \emph{Process A}, b) \emph{Process B}, and c) \emph{Process C}.  The cartoons depict corresponding MTL geometry, showing NbO\textsubscript{x} layers locations. d) EDS profiles of O/Nb content at the TEOS-SiO\textsubscript{2}/Nb interface corresponding to line-cuts indicated by the colored arrows in a), b) and c). The zero of the beam relative position corresponds to the TEOS-SiO\textsubscript{2}/Nb interface. 
We conjecture that the corresponding top \emph{or} bottom surfaces of a conducting strip (CS) and a ground plane (GP) have the same O/Nb profile. 
All STEM and EDS characterizations were done at EAG Laboratories.\cite{EAG}
\label{fig5}}
\end{figure*}

Figure~\ref{fig4}(b) shows that all three processes yield approximately the same loss tangent $\tan\! \delta \approx 1.2 \pm 0.1 \times 10^{-3}$ with virtually no MTL width dependence. This is in reasonable agreement with Kaiser, \cite{Kaiser2011} who observed $\tan\! \delta>3 \times 10^{-4}$ for sputtered amorphous SiO\textsubscript{2} at 4.2~K, 1-10~GHz. Therefore, in spite of the same CVD parameters, TEOS-SiO\textsubscript{2}  is not sensitive to significant differences between the three processes and offers a desirably large processing window.

\subsection{$R_{s0}$ variation with process}
To explain the Nb loss variation with process seen in Fig.~\ref{fig4}(a), we shall rely on a STEM complemented by an energy dispersive X-ray spectroscopy (EDS) with a 0.1-1 nm diameter electron beam and a detection limit of $>1\ at\%$.
Figure~\ref{fig5} shows STEM cross-sections of conducting strip looking into the 0.25 $\mu m$ wide MTL, for all three processes.  The analysis revealed that \emph{Processes A} and \emph{B} have 10-nm-thick Nb oxide (NbO\textsubscript{x}) layer\cite{Halbritter1987} on the top of both the ground plane and conducting strip, resulting from  the oxygen plasma treatment to promote the TEOS-SiO\textsubscript{2} adhesion. FIB cross-sectioning showed that NbO\textsubscript{x} has the same thickness in  both the ground plane and conducting strip, for all five MTL widths.
\emph{Process C} has no visible NbO\textsubscript{x} layers.

Figure~\ref{fig5} shows EDS line-cuts for the O/Nb content ratio at the top and bottom TEOS-SiO\textsubscript{2}/Nb interfaces of a conducting strip, for all three processes. At the time of EDS data collection, it was deemed that only the conducting strip is of interest, so the EDS mapping for the ground plane was not requested. 
However, using same fabrication recipe for all metal layers in a chip, we suppose that the corresponding top \emph{or} bottom surfaces of the conducting strip and ground plane have the same O/Nb profile, as conveyed in Fig.~\ref{fig5}(d). Comparison of O/Nb profiles between the $0.25\: \mu m$ and $1 \: \mu m$ conducting strips for \emph{Processes A} and \emph{C} suggests that for all three processes the profile is independent of MTL width.

Figure~\ref{fig3}(b) shows that for all MTL widths, the RF current concentrates in the ground plane and conducting strip near the TEOS-SiO\textsubscript{2}/Nb interfaces facing each other. Hence, the MTL resistive loss is mostly sensitive to the oxygen content at such interfaces. For \emph{Process A}, the oxygen content extends more than 10 nm into the top of the conducting strip, and diminishes within just few nanometers into the bottom of the ground plane. Conversely, for \emph{Process B}, the oxygen content extends more than 10 nm into the ground plane top, and diminishes within just few nanometers into the conducting strip bottom. At the same time, the conducting strip top in \emph{Process A} has a higher oxygen concentration than the ground plane top in \emph{Process B}. Moreover, the inverted microstrip geometry of \emph{Process A} makes NbO\textsubscript{x} layer to overlap with the current density peaks at the edges of the conducting strip. \cite{Sheen1991} 

These observations are consistent with about 2x difference in $R_s$ between \emph{Processes A} and \emph{B} for all MTL widths. 

\begin{figure*}[t!]
\includegraphics[width=\textwidth]{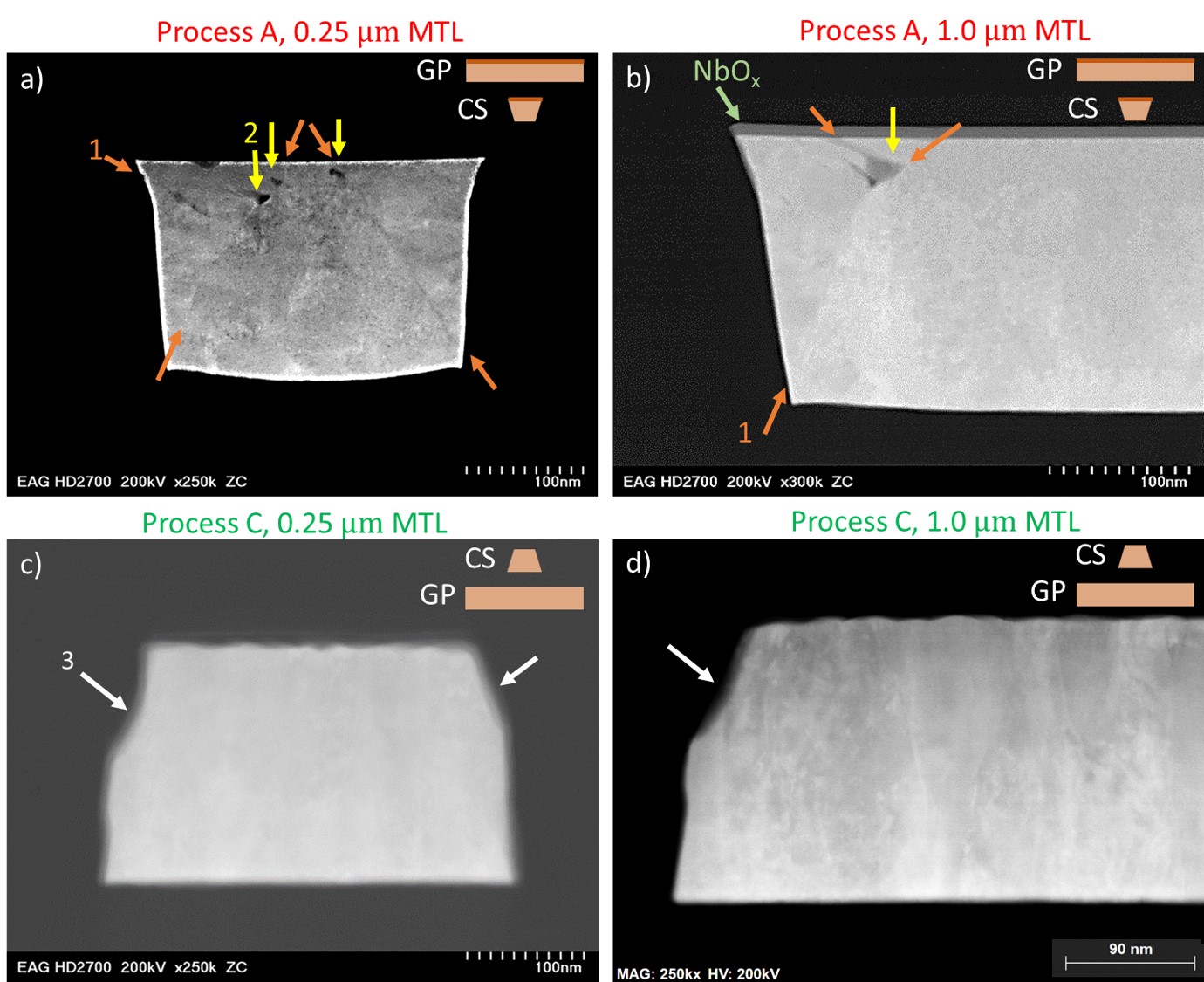}
\caption{STEM images taken for representative samples of 
a) $0.25 \:  \mu m$ and b) $1 \:  \mu m$ width MTL resonators from \emph{Process A}, 
c) $0.25 \:  \mu m$ and d) $1 \:  \mu m$ width MTL resonators from \emph{Process C}.
For reference, the MTL geometry is shown on the top right of every image. Image contrast in a), b), and d) was enhanced to highlight features. 
Orange arrows in a) and b) indicate the intersection between two Nb growth directions due to a damascene process.
Yellow arrows in a) and b) indicate voids and/or vacancies in Nb.
White arrows in c) and d) indicate the ``undercut" of the Nb during RIE.
All STEM and EDS characterizations were done at EAG Laboratories.\cite{EAG}
\label{fig6}}
\end{figure*}

For \emph{Process C}, the oxygen content diminishes within just few nanometers into both the ground plane top and the conducting strip bottom.  This is consistent with  \emph{Process C} exhibiting 30\% lower $R_{s}$ than \emph{Process B} for all MTL widths but $4 \:  \mu m$. We also note that the rough surface of the conducting strip in \emph{Process C} (see white arrows in Figs.~\ref{fig5}(c) and \ref{fig6}(c),(d)) caused by the RIE undercut does not affect the $R_s$ values. 

\subsection{$R_{s0}$ up-turn at $0.25 \: \mu m$ width}

Another data trend seen in Fig.~\ref{fig4}(a) is the 20-30\% up-turn in $R_{s}$ at 0.25 $\mu m$ MTL width for \emph{Processes A} and \emph{B}. STEM cross-sections in Fig.~\ref{fig6}(a),(b) show that the damascene process entails two Nb growth directions: vertical growth from the trench bottom and horizontal growth from the trench sidewall. \cite{Ju2002Trenchfill-sim} The two grain phases, a bottom grain and a sidewall grain, meet at about a 60{\textdegree} angle from the wafer plane. There are two such grain phase boundaries per conducting strip for both \emph{Processes A} and \emph{B}, indicated by orange arrows in Fig.~\ref{fig6}(a),(b). Conversely, Fig.~\ref{fig6}(c),(d) confirms that \emph{Process C} exhibits only a vertically grown grain.

Aligned with the current flow in conducting strip, the grain phase boundaries may not affect the current distribution. At the same time, morphology of the sidewall grain may reduce the electron mean free path $l_{mfp}$, increasing the BCS resistance $R_{BCS} \propto (1+\xi_0/l_{mfp})^{\frac{3}{2}} l_{mfp}$, where $\xi_0 \sim 38\ nm$ is the microscopic coherence length in pure Nb. Moreover, Nb voids and vacancies at the grain phase boundaries, indicated by yellow arrows in Fig.~\ref{fig6}(a),(b), may reduce $l_{mfp}$ and increase $R_{BCS}$. The voids can also act as Nb hydride formation sites,\cite{Romanenko2013} where hydrogen could diffuse into the Nb during fabrication, creating normal-conducting precipitates inside the conducting strip. Subject to proximity effect, hydrides suppress superconductivity in surrounding Nb. 

Since the sidewall grain geometry is determined by the trench depth, decreasing MTL width increases both the sidewall grain and void fractions in the conducting strip cross-sectional area. For instance, the sidewall grain fraction increases from about 20\% in a 1-$\mu m$-wide strip to about 50\% in a 0.25-$\mu m$-wide strip.  According to Fig.~\ref{fig3}(c) and Eq.~\ref{Eq. Qc}, the conducting strip contribution into the net $R_s$ shown in Fig.~\ref{fig4}(a) is the largest for a $0.25 \: \mu m$ MTL. Therefore, the $R_s$ up-turn for $0.25 \: \mu m$ MTL in \emph{Processes A} and \emph{B} can be attributed to the sidewall grain phase and/or voids present in conducting strip.

\section{Conclusions}

To conclude, we have developed a method to disentangle and quantify comparable superconductor and dielectric microwave losses by exploiting their frequency dependence in a multi-mode microstrip transmission line resonator representative of superconducting logic interconnects. The method was used to optimize a $0.25 \: \mu m$ planarized process for minimum interconnect loss. With the aid of the geometric factor concept and the 3D superconductor HFSS modeling, the intrinsic resistance $R_s$ was directly compared between different linewidths, stack geometries, and processing conditions. Correlating the Nb resistive loss with the STEM and EDS cross-sectional analysis revealed the mechanisms of loss above the microscopic theoretical minimum, including Nb oxide layer and Nb grain growth orientation.

We demonstrated that Nb interconnects can be scaled down to $0.25 \: \mu m$ linewidth with no penalty in microwave loss above the BCS minimum at 4.2 K. 
Nb sub-micron wires made by planarized Cloisonn\'{e} process exhibit resistive loss $R_s = 13 \pm 1.4 \: \mu \Omega$ at 4.2 K and 10 GHz, which is even lower than $R_s = 16-17\: \mu \Omega$ previously reported.\cite{SRIMP, Benvenuti1999, Taber1990}
We found that dielectric loss tangent $\tan\! \delta = 1.2 \pm 0.1 \times 10^{-3}$ for TEOS-derived SiO\textsubscript{2} remains unaffected by MTL geometries and processing conditions. This makes it a very attractive interconnect insulator for highly-integrated superconductor circuits, although the dielectric loss is fairly high. 

With Nb wires already at or below the theoretical minimum loss, it is worth exploring lower loss dielectrics compatible with the Nb fabrication.
The energy efficiency of a ZOR metamaterial clock network relative to the RQL logic at 4.2 K and 10 GHz can be improved from 30\% \cite{Strong2022} up to 80-90\% for Nb with $R_s = 13 \: \mu \Omega$ and a dielectric with $\tan\! \delta \sim 10^{-4}$.
Our loss data de-convolution method can be applied to any transmission line resonator including coplanar waveguide and stripline. 
We hypothesise that by increasing the resonator frequency range and number of modes, while improving the test probe bandwidth, one may unambiguously determine both the superconductor and dielectric loss frequency scaling powers in Eq.~\ref{Eq.7}, by allowing $\alpha$ and $p$ as the fitting parameters.






\section*{Acknowledgements}

The authors acknowledge Pavel Borodulin and Edward Kurek for assisting with test fixture design, Andrew Brownfield and David Vermillion for coordinating the test, Justin Goodman and Dr. Steve Sendelbach for assisting with data analysis, Dr. Eric Jones for the assistance with STEM and EDS analysis, Dr. Henry Luo for the penetration depth measurements by SQUID, and Dr. Flavio Griggio for the fruitful discussions. 
V.V.T. acknowledges insightful discussions with David Gill. S.M.A. acknowledges support from the National Science Foundation through Grant \#NSF DMR-2004386, and the U.S. Department of Energy/High Energy Physics through grant \#DESC0017931. 
This research is based on the work supported in part by the ODNI, IARPA, via ARO, contract \#W911NF-14-C-0116. 
The views and conclusions contained herein are those of the authors and should not be interpreted as necessarily representing the official policies or endorsements, either expressed or implied, of the NSF, DOE, ODNI, IARPA, or the US Government.
C.A.T.G. acknowledges approval for Public Release NG23-0122. © 2023 Northrop Grumman Systems Corporation

\appendix

\section{Derivation of Eq. \ref{EqQiGamma} \label{appendix a}}

Ignoring the radiation loss, the internal Q-factor of a transmission line resonator can be expressed as \cite{Pozar2011}
\begin{eqnarray}
\frac{1}{Q_{i}} = \frac{R}{\omega L} + \frac{G}{\omega C} 
\label{appendix A eq1}
\end{eqnarray}
where $R$ and $L$ are the line series resistance and inductance per unit length, $G$ and $C$ are the line shunt conductance and capacitance per unit length, and $\omega = 2 \pi f$ is the angular frequency with $f$ being the linear frequency. The telegrapher's equations have for the $R$, $L$, $G$, and $C$ of a superconducting transmission line\cite{Sass1968, Sheen1991, Pozar2011}
\begin{subequations}
\label{EqsRLGC}
\begin{gather}
R = 
\frac{2}{| I |^2} \int\displaylimits_{S}
R_s \lambda  | {\textbf J}|^2 {\rm d}s \approx \frac{2 R_{eff}}{w}  
\label{appendix A eq2a} \\[1ex]
L = \frac{\mu _0}{| I |^2} \int\displaylimits_{S} (| {\textbf H}|^2 + \lambda ^2 | {\textbf J}|^2)   {\rm d}s \approx \mu _0 \frac{s + 2 \lambda_{eff} }{w} 
\label{appendix A eq2b} \\[1ex]
G = \frac{\omega \varepsilon _0}{| V |^2} \int\displaylimits_{S} \tan\! \delta \: \varepsilon _r | {\textbf E}|^2  {\rm d}s \approx \omega \varepsilon _0 \varepsilon _r \frac{w}{s} \tan\! \delta 
\label{appendix A eq2c} \\[1ex]
C = \frac{\varepsilon _0}{| V |^2} \int\displaylimits_{S} \varepsilon _r | {\textbf E}|^2   {\rm d}s \approx \varepsilon _0 \varepsilon _r \frac{w}{s}
\label{appendix A eq2d}
\end{gather}
\end{subequations}
where the integrals are carried over the line cross-section $S$, $I$ and $V$ are the line current and voltage, $ {\textbf J}$, $ {\textbf H}$ and $ {\textbf E}$ are the vector current density, magnetic field and electric field, $R_s$ and $\lambda$ are the intrinsic resistance and magnetic penetration depth, $\varepsilon _0$ and $\mu _0$ are the vacuum permittivity and permeability, $\varepsilon _r$ and $\tan\! \delta$ are the relative dielectric constant and the loss tangent, and it is assumed that $\sigma _1 \ll \sigma _2$ in Eqs.~\ref{appendix A eq2a} and \ref{appendix A eq2b}. To provide a simple analytical reference, the approximations on the right of Eqs.~\ref{EqsRLGC} hold for a parallel-plate waveguide \cite{Taber1990, Talanov2000} formed by a dielectric spacer of thickness $s$ sandwiched between two superconducting plates of thickness $d$ and width $w \gg s$, with $R_{eff} = R_s [\coth(d/\lambda) + (d/\lambda) \csch ^2 (d/\lambda)]$ and $\lambda_{eff}=\lambda \coth (d/\lambda)$ being the plate effective surface resistance and effective penetration depth, \cite{Klein1990} respectively.

Consider a uniform transmission line formed by $M$ conductors, and $N$ dielectric layers or tubes. Inserting Eqs.~\ref{EqsRLGC}
into Eq.~\ref{appendix A eq1} yields 
\begin{eqnarray}
\frac{1}{Q_i}= \sum_{m=1}^M \frac{\overline{R_{sm}}}{\Gamma _{cm}} +  \sum_{n=1}^N \frac{\overline{\tan \!\delta _{n}}}{\Gamma _{dn}} 
\label{appendix A eq3}
\end{eqnarray}
 Here $\overline{R_{sm}}$ and $\overline{\tan\! \delta _n}$ are the averaged quantities describing loss in the $m$-th conductor and $n$-th dielectric,
\begin{subequations}
\label{RsTand}
\begin{gather}
\overline{R_{sm}} = \frac{\int_{S_{cm}} R_s \lambda |\textbf J|^2 {\rm d}s}{ \int   _{S_{cm}} \lambda  |\textbf J|^2  {\rm d}s} 
\label{RsTandA} \\[1ex]
\overline{\tan \!\delta_n} = \frac{\int  _{S_{dn}} \tan\!\delta \: \varepsilon _r |\textbf E|^2  {\rm d}s}{\int   _{S_{dn}} \varepsilon _r |\textbf E|^2 {\rm d}s} 
\label{RsTandB}
\end{gather}
\end{subequations}
where $S_{cm}$ and $S_{dn}$ are the cross-sectional areas of the $m$-th conductor and $n$-th dielectric. In the case of the homogeneous losses (with arbitrary $\lambda$ and $\varepsilon_r$ distributions), $\overline{R_{sm}}=R_{sm}$ and $\overline{\tan\! \delta_n} = \tan\! \delta_n$. Furthermore, in Eq.~\ref{appendix A eq3}, the partial geometric factors $\Gamma _{cm}$ and $\Gamma _{dn}$ associated with respective losses in the $m$-th conductor and $n$-th dielectric are
\begin{subequations}
\label{GeomFact}
\begin{gather}
\Gamma _{cm}= \omega \mu _0 \frac{\int   _{S} (|\textbf H|^2 + \lambda ^2 |\textbf J|^2) {\rm d}s}{ 2 \int   _{S_{cm}} \lambda  |\textbf J|^2  {\rm d}s} 
\label{appendix A eq4a} \\[1ex]
\Gamma _{dn}=\frac{\int   _{S} \varepsilon _r |\textbf E|^2  {\rm d}s}{\int   _{S_{dn}} \varepsilon _r |\textbf E|^2 {\rm d}s} 
\label{appendix A eq4b}
\end{gather}
\end{subequations}

The conductor geometric factor $\Gamma _{c}$ of a superconducting transmission line resonator has units of Ohm, and is defined exclusively by the line cross-sectional geometry and penetration depth $\lambda$. A good superconductor with $\sigma_1 \ll \sigma_2$ makes $\lambda$ frequency independent, and so the fraction on the right of Eq.~\ref{appendix A eq4a}. Note that definition~\ref{appendix A eq4a} involves the field and current density inside the superconducting members. This differs from a cavity geometric factor\cite{Turneaure1991, Krupka1998, Hein1999}
\begin{eqnarray}
\label{GamCav}
\Gamma _{cav}= \omega \mu _0 \frac{\int_V |\textbf H|^2 {\rm d}v}{\int_A |\textbf H_{\tau}|^2  {\rm d}a} = \omega \mu _0 D
\end{eqnarray}
which is governed by the ratio of the cavity volume $V$ to the walls area $A$ and involves magnetic field within that volume only, giving for the cavity Q-factor $Q = \Gamma _{cav}/R_s$. Equation~\ref{GamCav} is applicable in the case of $D \gg \lambda$, where $D$ is on the order of the cavity smallest dimension.\cite{VainshteinBook}  The concept of a cavity geometric factor is associated with Leontovich's impedance boundary condition $\textbf{E}_{\tau}=Z_s \textbf{H}_{\tau} \cross \textbf{n}$, where $\textbf{E}_{\tau}$ and $\textbf{H}_{\tau}$ are the respective tangential electric and magnetic fields at the impedance surface, and $\textbf{n}$ is the inward unit vector normal to the surface.\cite{Leontovich1944, Leontovich1948, Miller1961, senior1960impedance} The quantity $\Gamma_c^{-1}$ can be seen as a transmission line counterpart of a conductor participation ratio $\Gamma _{cav}^{-1}$ found in voluminous, cavity-like resonators.\cite{Mcrae2020}

The dielectric geometric factor of a transmission line resonator given by Eq.~\ref{appendix A eq4b} is unitless, and is defined exclusively by the line cross-sectional geometry and $\varepsilon _r$. The quantity $\Gamma_d^{-1}$ can be seen as a transmission line counterpart of a dielectric filling factor found in the voluminous resonators\cite{Krupka1998, Mcrae2020}
\begin{subequations}
\label{GeomFact2}
\begin{gather}
\nonumber
p_{dn} = \frac{\int   _{V_{n}} \varepsilon _r |\textbf E|^2 {\rm d}v}{\int   _{V} \varepsilon _r |\textbf E|^2  {\rm d}v} 
\end{gather}
\end{subequations}
where $V_n$ is the volume of the $n$-th dielectric ($n \ge 1$), and $V$ is the volume of the entire resonator.

An embedded MTL like in Fig. ~\ref{fig1}(b) calls for $M=2$ and $N=2$ or 3 in Eq.~\ref{appendix A eq3}, depending on the process. Assuming  homogeneous losses within each of the conductor or dielectric members, Eq.~\ref{appendix A eq3} gives rise to
\begin{subequations}
\label{AppEqQiGamma}
\begin{gather}
\frac{1}{Q_i^A} =  \frac{R_{s}^{gp}}{\Gamma_c^{gp}} + \frac{R_{s}^{cs}}{\Gamma_c^{cs}} + \frac{\tan \! \delta ^{SiO2}}{\Gamma_d^{SiO2}} + \frac{\tan \! \delta ^{Si}}{\Gamma_d^{Si}} \label{AppEqQiGammaa} \\[1ex]
\frac{1}{Q_i^{B,C}} =  \frac{R_{s}^{gp}}{\Gamma_c^{gp}} + \frac{R_{s}^{cs}}{\Gamma_c^{cs}} + \frac{\tan \! \delta ^{SiO2}}{\Gamma_d^{SiO2}} + \frac{\tan \! \delta ^{SiN}}{\Gamma_d^{SiN}} + \frac{\tan \! \delta ^{LHe}}{\Gamma_d^{LHe}} \label{AppEqQiGammab}
\end{gather}
\end{subequations}
where Eqs.~\ref{AppEqQiGammaa} and \ref{AppEqQiGammab} correspond to \emph{Process A} and \emph{Processes B, C}, respectively. Furthermore, $R_{s}^{gp}$ and $R_{s}^{cs}$ are the intrinsic resistances of the ground plane and conducting strip, $\Gamma_c^{gp}$ and $\Gamma_c^{cs}$ are the partial geometric factors associated with resistive loss in respective conductors, $\tan \! \delta ^{SiO2}$, $\tan \! \delta ^{Si}$, $\tan \! \delta ^{SiN}$ and $\tan \! \delta ^{LHe}$ are the dielectric loss tangents of the TEOS-SiO\textsubscript{2} insulator, Si substrate, SiN\textsubscript{x} passivation layer and LHe, respectively, and $\Gamma_d^{SiO2}$, $\Gamma_d^{Si}$, $\Gamma_d^{SiN}$ and  $\Gamma_d^{LHe}$ are the partial geometric factors associated with loss in respective materials. Due to $\tan \! \delta ^{SiN} \ll \tan \! \delta ^{SiO2}, \tan \! \delta ^{Si}, \tan \! \delta ^{LHe}$ \cite{Kaiser2011}, and $\Gamma_d^{SiO2} \ll \Gamma_d^{SiN}, \Gamma_d^{Si}, \Gamma_d^{LHe}$, the Si, SiN\textsubscript{x} and LHe loss contributions can be ignored in Eqs.~\ref{AppEqQiGamma}, both leading to Eq.~\ref{EqQiGamma}.



\section{Derivation of Eqs.~\ref{EqsZY}} \label{appendix b}

Consider a two-port network formed by a transmission line of length $l$. The ABCD (transmission) matrix of such network is \cite{paul2007analysis}
\begin{eqnarray}
\begin{bmatrix}
A & B \\
C & D
\end{bmatrix}
=
\begin{bmatrix}
\cosh(\gamma l) & Z_{TL} \sinh(\gamma l)  \\ Z_{TL} ^{-1} \sinh(\gamma l) & \cosh(\gamma l) 
\end{bmatrix}
\label{appendix B eq1}
\end{eqnarray}
where $\gamma$ is the propagation constant, and $Z_{TL}$ is the characteristic impedance. For the electrically short network, a quadratic Taylor expansion around $\gamma l = 0$ yields $\cosh (\gamma l) \approx 1 + (\gamma l)^2 / 2$ and $\sinh(\gamma l) \approx \gamma l$. A general transmission line has \cite{Ramo1994} $\gamma = \sqrt{ZY}$ and $Z_{TL} = \sqrt{Z/Y}$, where $Z$ and $Y$ are the series impedance and shunt admittance per unit length. Inserting all of the above into Eq.~\ref{appendix B eq1} gives
\begin{eqnarray}
\begin{bmatrix}
A & B \\
C & D
\end{bmatrix}
\approx
\begin{bmatrix}
1 + Z Y l^2 / 2 & Z l \\  Y l & 1 + Z Y l^2 / 2 
\end{bmatrix}
\label{appendix B eq2}
\end{eqnarray}
By reciprocity, the elements of the [Z]-matrix corresponding to the ABCD matrix\cite{Pozar2011} given by Eq.~\ref{appendix B eq2} are $Z_{11} = Z_{22} = (Y l)^{-1} + Z l /2$ and $Z_{12}=Z_{21}=(Y l)^{-1}$. Solving these for the $Z$ and $Y$ yields Eqs.~\ref{EqsZY}.
%
%
%
%

\clearpage


\bibliography{0-final-draft}
\end{document}